\documentclass[12pt]{article}

\usepackage{mathptmx}
\usepackage{txfonts}
\usepackage{xcolor}

\usepackage{graphicx}

\usepackage[letterpaper,margin=1in]{geometry}

\renewenvironment{abstract}
	{\quotation}
	{\endquotation}

\date{}

\makeatletter
\renewcommand{\fnum@figure}{\textbf{Figure \thefigure}}
\renewcommand{\fnum@table}{\textbf{Table \thetable}}
\makeatother

\usepackage{scicite}

\usepackage{url}
\usepackage{lineno}
\def\scititle{
	Spontaneous emission of light by non-equilibrium phonons
}
\title{\bfseries \boldmath \scititle}

\author{
	J. Plo$^{1\dagger}$,
	P. Valvin$^{1\dagger}$,
	M. Moret$^{1}$,
	T. Taliercio$^{2}$,
	J. Batista$^{1}$,\and
	T. Sohier$^{1}$,
	A. Vasanelli$^{3}$,
	C. Sirtori$^{3}$,
	B. Gil$^{1}$,
	W. Desrat$^{1}$,
	G. Cassabois$^{1,4\ast}$\and
	\small$^{1}$Laboratoire~Charles~Coulomb, CNRS-Universit\'e~de~Montpellier, 34095 Montpellier, France.\and
	\small$^{2}$Institut d'Electronique et des Systèmes, CNRS-Universit\'e~de~Montpellier, 34095 Montpellier, France.\and
	\small$^{3}$Laboratoire de Physique de l’Ecole
Normale Supérieure, 75005 Paris, France.\and
	\small$^{4}$Institut Universitaire de France, 75231 Paris, France.\and
	\small$^\ast$Corresponding author. Email: guillaume.cassabois@umontpellier.fr\and
	\small$^\dagger$These authors contributed equally to this work.
}

\begin{document} 

% Insert the title and author list
\maketitle

% Abstract, in bold
% There are strict length limits, and not all formats have abstracts.
% Consult the journal instructions to authors for details.
% Do not cite any references in the abstract.
\begin{abstract} \bfseries \boldmath
% Start with one or two sentences of background
When a system is brought out of equilibrium by an external excitation, its relaxation to thermodynamic equilibrium generates phonons. These non-equili\-brium phonons degrade via cascaded anharmonic decay processes, progressively leading to a thermal population of phonons following a Bose-Einstein distribution at the system temperature. Preceding heat dissipation by convection, conduction and incandescence, this early phase of the relaxation dynamics is commonly assumed to be exclusively non-radiative. Here, we demonstrate that the radiative emission by phonons can be an efficient relaxation pathway competing with the intrinsic anharmonic decay. Optical spectroscopy under femtosecond two-photon excitation in boron nitride unveils a photoluminescence signal in the mid-infrared spectral range, stemming from the spontaneous emission of light by non-equilibrium phonons. This observation of non-thermal radiation from phonons introduces a new paradigm for out-of-equilibrium physics, mid-infrared optics, and thermal management.
\end{abstract}

% The first paragraph of any Science paper does NOT have a heading
% Nor is it indented
\noindent

Semiconductor optoelectronics relies on non-equilibrium charge carriers that are generated under optical or electrical excitation. A universal detrimental effect is the non-radiative recombination of charge carriers at trapping centers \cite{sr,h}, which can be drastically reduced by improving the quality of the semiconductor crystals, as routinely performed today in any device technology. In contrast to non-equilibrium charge carriers, the relaxation dynamics of non-equilibrium phonons is driven by intrinsic non-radiative processes of high efficiency. They stem from the anharmonicity of the lattice forces, which makes a phonon decay into two or three phonons of lower energies. The phonon lifetime is thus mainly determined by the anharmonic decay time, which strongly depends on the phonon mode. The higher the energy, the higher the density of final states, so that low-energy acoustic phonons can last for seconds, while high-energy optical phonons have lifetimes typically lying in the picosecond range \cite{Bro85,Mac20}. Anharmonicity inherently imposes a stringent constraint on the radiative lifetime of optical phonons, which generally results in negligible radiative efficiencies for phonons.

In this context, boron nitride (BN) has been predicted to be an outstanding platform to investigate the spontaneous emission of light by non-equilibrium phonons \cite{Cas22PRX,Cas22PRR}. Because BN is a polar crystal made of light elements, the photon-phonon interaction reaches an exceptional magnitude, with a radiative lifetime of few picoseconds, as estimated from reflectivity measurements in atomically-thin BN \cite{Cas22PRX}. The BN advantage is expected to extend to multilayer crystals of sub-wavelength thickness thanks to the superradiance of optical phonons at the crossover from weak to strong coupling \cite{Cas22PRR,Ma22}.

Our experimental evidence for spontaneous emission of light by non-equilibrium phonons relies on optical spectroscopy under femtosecond two-photon excitation of BN crystals at cryogenic temperatures ($T$$\sim$25 K). The cryogenic environment ensures a spectral separation between the black-body emission of the sample ($\lambda$$\sim$115 $\mu$m at 25 K) and the photoluminescence (PL) of the photo-generated phonons lying in the mid-infrared (MIR) domain ($\lambda$$\sim$7 $\mu$m in BN). The femtosecond laser pulses ($\lambda$$\sim$400 nm) realize an ultrafast two-photon excitation on a timescale much shorter than the carrier and phonon relaxation dynamics \cite{CassaboisNP}. The temporal spacing of $\sim$12 ns between two consecutive pulses is one decade larger than the electronic and thermal relaxation times \cite{wat11,casPRB16,kan21,sha23,sca25,hut25}, thus avoiding any quasi-continuous driving of the system and ensuring periodic ultrafast excitation. Insightful access to the spatial variations of the MIR emission of phonons is provided by means of a home-made cryogenic scanning confocal microscope, with an achromatic design allowing ultrabroadband optical spectroscopy with a spatial resolution given by the diffraction limit until the UV-C spectral range ($\lambda$$\sim$200 nm) \cite{Val19,Rou21,Rou22,Moo25}. Layered BN with sp$^2$ hybridization of the atomic orbitals cristallizes in various polytypes differing by their stacking order \cite{gil}, the most stable polytypes being hexagonal BN (h-BN) \cite{wat04} and rhombohedral BN (r-BN) \cite{sato}. The non-thermal radiation from phonons is unveiled here by experiments in r-BN with state-of-the-art optical properties \cite{des25}.
%%%%%%%%%%%%%%%%%%%%%%%%%%%%%%%%%%%%%%%%%%%%%%%%%%%%%%%%%%%%%%
\begin{figure} % Do NOT use \begin{figure*}
	\centering
	\includegraphics[width=0.65\textwidth]{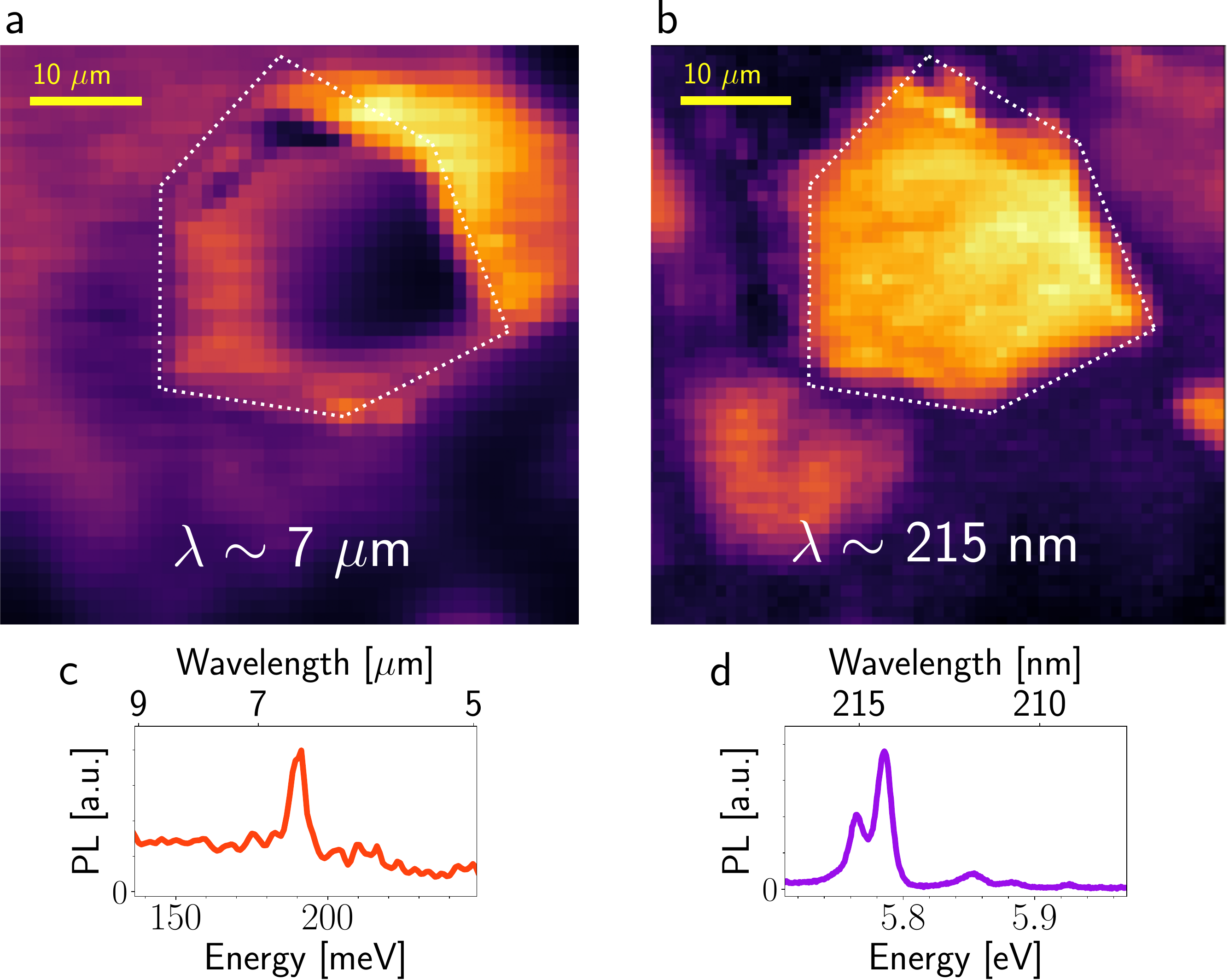} % for an image file named example_figure.*
	% Pick an appropriate width - in print, figures are usually one or two columns wide, which can
	% be approximated by 0.3\textwidth or 0.6\textwidth respectively. Use appropriate label sizes.

	% Captions go below figures
	\caption{\textbf{IR-UV dual-band imaging.} Photoluminescence (PL) spectroscopy in rhombohedral boron nitride under two-photon excitation at $\lambda$=397 nm, for a sample temperature $T_s$=25 K: (a) map of the phononic PL lying in the MIR range at $\lambda$$\sim$7 $\mu$m (linear color map), (b) map of the excitonic PL lying in the UV-C range at $\lambda$$\sim$215 nm (logarithmic color map), and corresponding PL spectra in (c) and (d), respectively. The dotted lines follow the microcrystallite edges as determined by electron microscopy (see Fig.S1).}
	\label{figmaps} % give each figure a logical label name
\end{figure}
%%%%%%%%%%%%%%%%%%%%%%%%%%%%%%%%%%%%%%%%%%%%%%%%%%%%%%%%%%%%%%
\subsection*{Infrared-Ultraviolet dual-band imaging}
Imaging of the MIR emission under two-photon excitation is performed in polycrystalline r-BN composed of high-quality microcrystallites of parallel c-axis but of various lateral sizes and thicknesses (see details in Methods). Fig.\ref{figmaps}a is an example in a $\sim$50$\times$50 $\mu$m$^2$ region containing a microcrystallite of hexagonal shape, where the dotted lines follow the microcrystallite edges as determined by electron microscopy (see Fig.S1). The spatial pattern of the MIR emission comprises a weak and slowly varying background, together with the luminescing hexagon with a faint core and a bright outer corner corresponding to the microcrystallite sidewall (as seen by electron microscopy in Fig.S1). The emission spectrum shows a prominent line at 193 meV [Fig.\ref{figmaps}c], lying in between the transverse optical (TO) and longitudinal optical (LO) phonons of energy 170 and 200 meV in bulk BN, respectively \cite{des25,cal19}. The present MIR signal corresponds to the first detection of PL from optical phonons, not only in BN but in any solid-state or molecular system.

Our scanning cryomicroscope with an achromatic design allows dual-band imaging in the exact same region, so that the PL intensity in the UV-C range can be mapped with a comparable spatial resolution as in the MIR domain. The spatial distribution of the emission at the edge of the 6 eV-electronic bandgap of r-BN is shown in Fig.\ref{figmaps}b, together with the PL spectrum in Fig.\ref{figmaps}d consisting in the series of phonon replicas of the indirect exciton at $\sim$5.95 eV \cite{des25,vuo17}. While the excitonic PL is spatially homogeneous within the r-BN microcrystallite, the phononic PL is more intense at the edges of the hexagonal surface and at the tilted sidewall located around its top-right corner. This observation suggests a complex angular dependence of the phononic PL, which appears to better match the solid angle of collection of our reflective microscope objective (see details in Methods) at specific locations.

Remarkably, our spatially-resolved PL measurements unveil the variations of the MIR signal at a spatial scale of $\sim$400 nm, which is well beyond the wavelength of the MIR emission although our experiments are performed in far-field. The $\sim$$\lambda$/20 spatial resolution in the MIR map detected at $\lambda$$\sim$7 $\mu$m in Fig.\ref{figmaps}a results from the considerable spectral detuning bewteen detection and excitation in our ultrabroadband configuration of PL spectroscopy, where the spatial resolution is essentially determined by the diffraction-limited excitation spot (see details in Supplementary Information).
%%%%%%%%%%%%%%%%%%%%%%%%%%%%%%%%%%%%%%%%%%%%%%%%%%%%%%%%%%%%%%
\begin{figure} % Do NOT use \begin{figure*}
	\centering
	\includegraphics[width=0.65\textwidth]{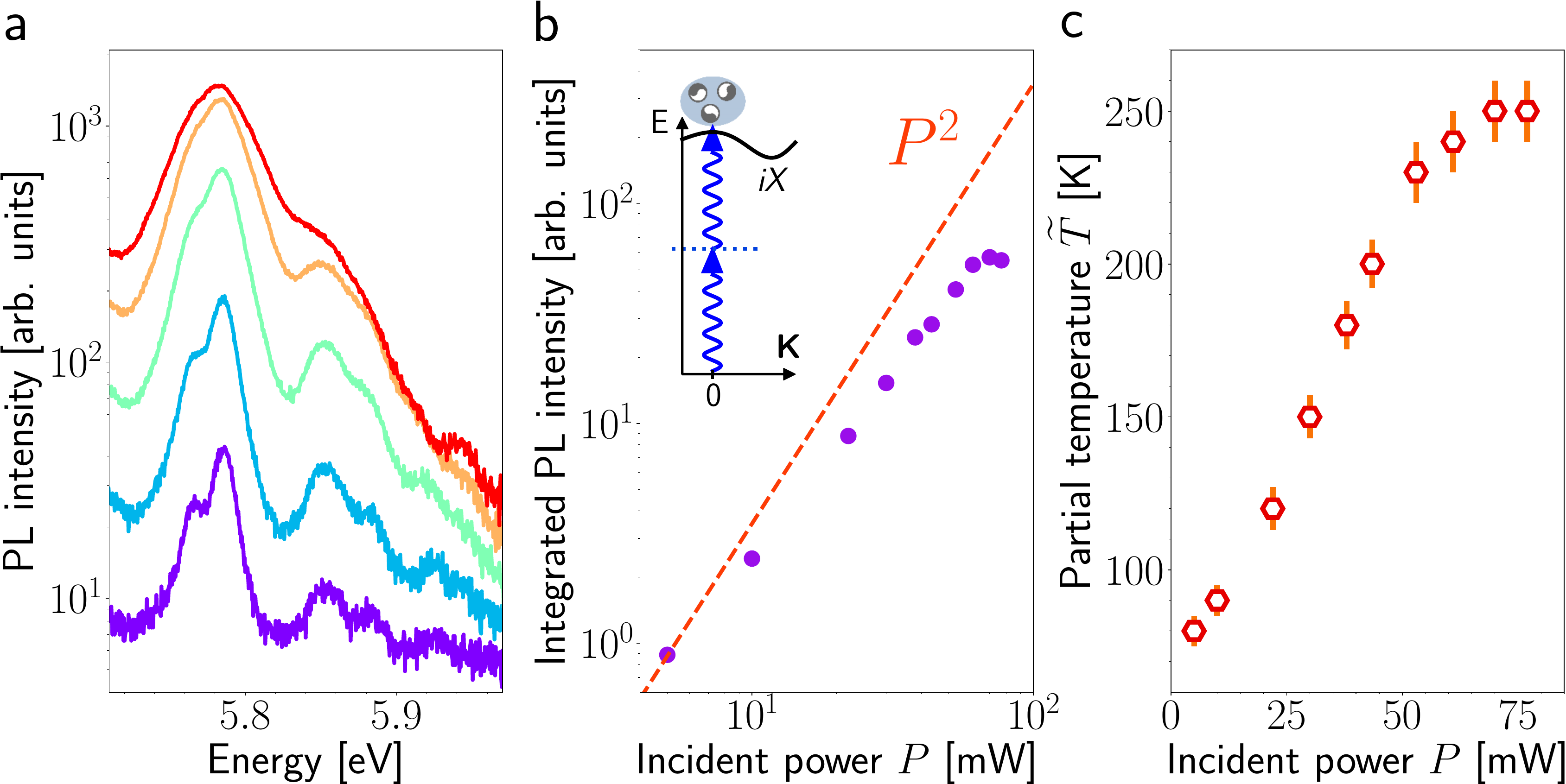} % for an image file named example_figure.*
	% Pick an appropriate width - in print, figures are usually one or two columns wide, which can
	% be approximated by 0.3\textwidth or 0.6\textwidth respectively. Use appropriate label sizes.

	% Captions go below figures
	\caption{\textbf{Excitonic photoluminescence.} Power-dependent two-photon excitation of the excitonic photoluminescence in the UV-C range: (a) spectra for excitation powers of 5, 10, 22, 38 and 60 mW (from bottom to top), and (b) integrated intensity versus excitation power. Inset: schematic of excitons generation at \textbf{K}$\sim$0 by two-photon absorption in indirect-gap bulk BN, where $iX$ stands for the indirect fundamental exciton, giving rise to the phonon-assisted emission displayed in panel (a). (c) Excitation-power dependence of the partial temperature $\widetilde{T}$ of the non-equilibrium phonons controlling the excitonic broadening. Excitation wavelength $\lambda$=397 nm, sample temperature $T_s$=25 K.}
	\label{figPLUV} % give each figure a logical label name
\end{figure}
%%%%%%%%%%%%%%%%%%%%%%%%%%%%%%%%%%%%%%%%%%%%%%%%%%%%%%%%%%%%%%
%%%%%%%%%%%%%%%%%%%%%%%%%%%%%%%%%%%%%%%%%%%%%%%%%%%%%%%%%%%%%%
\subsection*{Indirect Auger processes}
The radiative decay of non-equilibrium optical phonons presented in Fig.\ref{figmaps} follows their photo-generation during excitonic collisions. The phonons do not originate from the phonon-assisted recombination of the fundamental indirect exciton [labelled $iX$, inset of Fig.\ref{figPLUV}b] in BN because of the large associated momentum incompatible with direct spontaneous emission of light \cite{CassaboisNP,vuong,des25}. Strong Coulomb interactions in BN manifest by the large excitonic binding energy on the one hand \cite{CassaboisNP,sch19}, and by the efficient exciton-exciton collisions on the other hand \cite{pla19,sha23,sca25}. These many-body effects include exciton-exciton annihilation, contributing to the non-radiative relaxation of the population, and elastic exciton-exciton scattering, redistributing excitons within bands and inducing pure dephasing. Scattering by phonons widens the microscopic configurations of the Auger mechanisms, and such indirect processes play a dominant role in wide-bandgap semiconductors such as III-nitrides \cite{kio13}. Here, phonon-assisted excitonic nonlinearities drive the generation of the phonons that radiatively decay in the MIR.

Exciton-exciton collisions are first apparent in excitation-power dependent measurements of the PL detected in the UV-C range [Fig.\ref{figPLUV}]. For increased intensities of the laser beam at $\lambda$$\sim$400 nm, the PL signal rises but at a slower pace than the quadratic dependence expected under two-photon absorption, as shown in Fig.\ref{figPLUV}a,b. This phenomenology evidences efficient Auger processes, in agreement with previous reports on the perturbation of the excitonic relaxation dynamics by exciton-exciton annihilation \cite{pla19,sha23,sca25}, and consistently with the rate equation model detailed in the Supplementary Information. Simultaneously, there is a broadening of all emission lines in the series of phonon replicas composing the PL spectrum [Fig.\ref{figPLUV}a]. This excitation-induced broadening can be of two origins: either exciton-exciton collisions, or exciton-phonon scattering. In low-dimensional systems with comparably large excitonic binding energies such as atomically-thin 2D crystals and 1D single-wall carbon nanotubes, the influence of exciton-exciton annihilation is ubiquitous in the population relaxation dynamics \cite{wang04,sun14}, while the dephasing due to exciton-exciton collisions is at most in the range of few meVs \cite{ngu11,moo15}, i.e. one order of magnitude lower than the few tens of meV observed in Fig.\ref{figPLUV}a. The excitation-induced broadening observed here can alternatively come from the emission of phonons during the non-radiative relaxation of excitons. In semiconductors, the thermally-assisted broadening is typically mediated by acoustic phonons below 50 K, and by optical phonons at higher temperatures. By temperature-dependent experiments in isotopically-purified crystals, it was demonstrated that the specific optical phonons dominating the thermally-assisted broadening in layered BN have a characteristic energy of $\sim$15 meV \cite{vuong,vuongPRB}.

In order to quantify the population of this sub-class of vibrational modes at a given excitation power, we introduce their partial temperature $\widetilde{T}$, in analogy with the partial pressures of the constituents inside a mixture. The partial temperature $\widetilde{T}$ at an excitation power $P$ is defined as the sample temperature $T_s$ leading to the same excitonic broadening as measured in temperature-dependent PL experiments at low excitation power (see Fig.S2). In other words, for the specific lattice vibrations responsible of the excitonic broadening, their non-equilibrium population at an excitation power $P$ matches a thermal Bose-Einstein distribution at $\widetilde{T}$. This definition implicitly assumes that the excitonic collisional broadening is negligible, so that the estimated $\widetilde{T}$ provides, strictly speaking, an upper bound for the photo-generation of these non-equilibrium phonons. The variations of $\widetilde{T}$ over the investigated range of excitation power are displayed in Fig.\ref{figPLUV}c. The steep increase of $\widetilde{T}$ from 80 to 250 K suggests the efficient generation of non-equilibrium phonons by indirect Auger processes, at least for the particular sub-class of lattice vibrations controlling the excitonic broadening.
%%%%%%%%%%%%%%%%%%%%%%%%%%%%%%%%%%%%%%%%%%%%%%%%%%%%%%%%%%%%%%
\subsection*{Ruling out photo-thermal effects}
At this point, one may wonder if the strong $\widetilde{T}$ variations in Fig.\ref{figPLUV}c do not simply come from a local heating of the BN crystal under photo-excitation, the so-called partial temperature $\widetilde{T}$ being nothing else than the effective sample temperature $T_s$ under the laser spot. This is a critical question. If it would be the case, the detected MIR emission at $\lambda$$\sim$7 $\mu$m in Fig.\ref{figmaps} would not arise from the spontaneous emission of photons by non-equilibrium phonons, but it would stem from the incandescence of the BN crystal locally heated by the excitation laser. In order to test this scenario, we performed the control experiment consisting in studying the emission of our sample at $T_s$=250 K, without any optical excitation ($P$=0) (see details in Methods). No MIR signal was detected. This rules out any trivial origin of the MIR emission shown in Fig.\ref{figmaps} as resulting from a photo-thermal effect. Conversely, it proves the non-equilibrium nature of the distribution of phonons in our BN crystals under photo-excitation, and consequently the non-thermal origin of the detected MIR emission in Fig.\ref{figmaps}, which can be unambiguously attributed to the PL of phonons.
%%%%%%%%%%%%%%%%%%%%%%%%%%%%%%%%%%%%%%%%%%%%%%%%%%%%%%%%%%%%%%
\begin{figure} % Do NOT use \begin{figure*}
	\centering
	\includegraphics[width=0.65\textwidth]{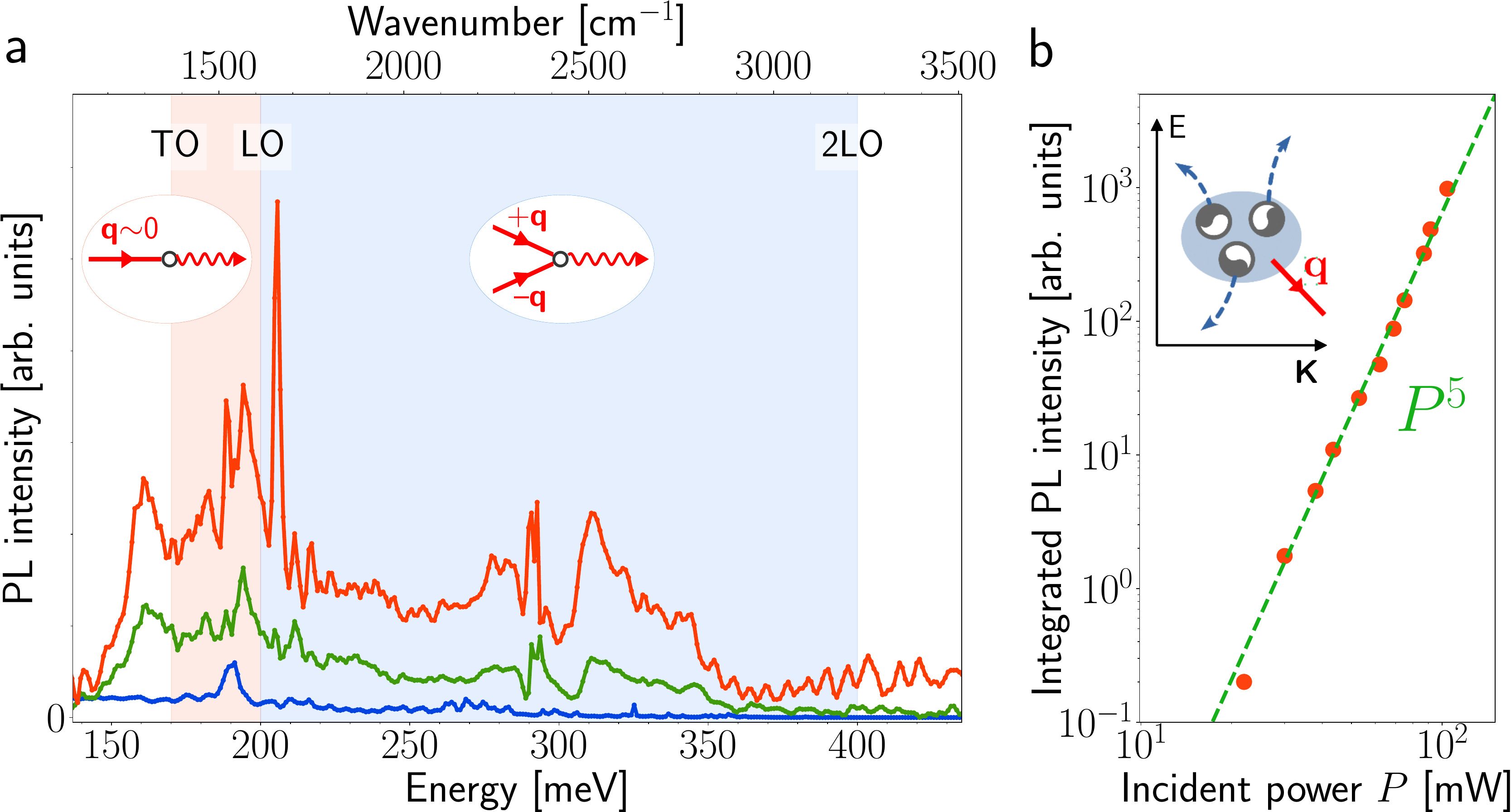} % for an image file named example_figure.*
	% Pick an appropriate width - in print, figures are usually one or two columns wide, which can
	% be approximated by 0.3\textwidth or 0.6\textwidth respectively. Use appropriate label sizes.

	% Captions go below figures
	\caption{\textbf{Phononic photoluminescence.} Power-dependent two-photon excitation of the phononic photoluminescence in the MIR range: (a) spectra for $P_0$, 1.5$P_0$ and 2$P_0$ (from bottom to top, with $P_0$=45 mW), and (b) integrated intensity versus excitation power. Insets: (a) diagrams of one-phonon (\textbf{q}$\sim$0) and two-phonon (of opposite $+$\textbf{q} and $-$\textbf{q} wavevectors) radiative decay processes, and (b) schematic of three-exciton collisions assisted by phonon emission accounting for the $P^5$ dependence of the phononic PL. Excitation wavelength $\lambda$=397 nm, sample temperature $T_s$=25 K.}
	\label{figPLMIR} % give each figure a logical label name
\end{figure}
%%%%%%%%%%%%%%%%%%%%%%%%%%%%%%%%%%%%%%%%%%%%%%%%%%%%%%%%%%%%%%
%%%%%%%%%%%%%%%%%%%%%%%%%%%%%%%%%%%%%%%%%%%%%%%%%%%%%%%%%%%%%%
\subsection*{Emission inside and beyond the reststrahlen band}
The phonon-assisted exciton-exciton collisions inferred from PL spectroscopy in the UV-C range [Fig.\ref{figPLUV}] are conspicuous in the excitation-power dependence of the phononic PL [Fig.\ref{figPLMIR}]. The MIR emission rises extremely abruptly with the intensity of the laser beam at $\lambda$$\sim$400 nm [Fig.\ref{figPLMIR}a], following a power law with an exponent as high as $\sim$5 [Fig.\ref{figPLMIR}b]. This ultrastrong nonlinearity surpasses the one recorded for the excitonic PL under two-photon excitation [Fig.\ref{figPLUV}b]. It reveals the complexity of the many-body effects leading to the generation of phonons that radiatively decay in the MIR range in BN. As detailed in the Supplementary Information, three-exciton collisions assisted by phonon emission [Fig.\ref{figPLMIR}b, inset] account for the $P^5$ dependence of the phononic PL.

Notably, the MIR PL spectrum does not only consist in a narrow line at 193 meV, in between the TO and LO phonons delimitating the reststrahlen band of bulk BN [pink shaded area in Fig.\ref{figPLMIR}a], but also in a broad emission band extending up to twice the LO energy [blue shaded area in Fig.\ref{figPLMIR}a], which becomes increasingly visible at higher excitation power. More specifically, the fraction of the PL intensity above the LO phonon energy smoothly increases by a factor $\sim$2 in the investigated range of excitation power. Infrared absorption above the reststrahlen band comes from multiphonon absorption due to anharmonicity \cite{Car96}. These higher-order processes lead to weak absorption coefficients that exponentially decrease with the number of involved phonons \cite{spa73}, which wavevectors are no longer restricted to the zone center but can span the full reciprocal phase, provided their sum lies inside the light cone. By time-reversal symmetry, multiphonon radiative relaxation yields additional decay channels in the non-equilibrium phonon dynamics. The growing contribution of two-phonon radiative decay to the phononic PL directly reflects the activation of this relaxation channel by the population of non-equilibrium phonons, as explained by our theoretical model presented in the Supplementary Information. From the minor impact at low $P$ where the PL is dominated by one-phonon radiative decay, two-phonon radiative decay has a rate progressively increasing as a function of the excitation power, the PL intensity fraction stemming from two-phonon processes being enhanced by a factor $\sim$2 in our measurements, until two-phonon radiative decay eventually becomes the major part of the phononic PL spectrum [Fig.\ref{figPLMIR}a]. Such a smoothly changing efficiency balance between one- and two-phonon radiative decays appears as a key phenomenological signature for the spontaneous emission of light by non-equilibrium phonons.
%%%%%%%%%%%%%%%%%%%%%%%%%%%%%%%%%%%%%%%%%%%%%%%%%%%%%%%%%%%%%%
\begin{figure} % Do NOT use \begin{figure*}
	\centering
	\includegraphics[width=0.65\textwidth]{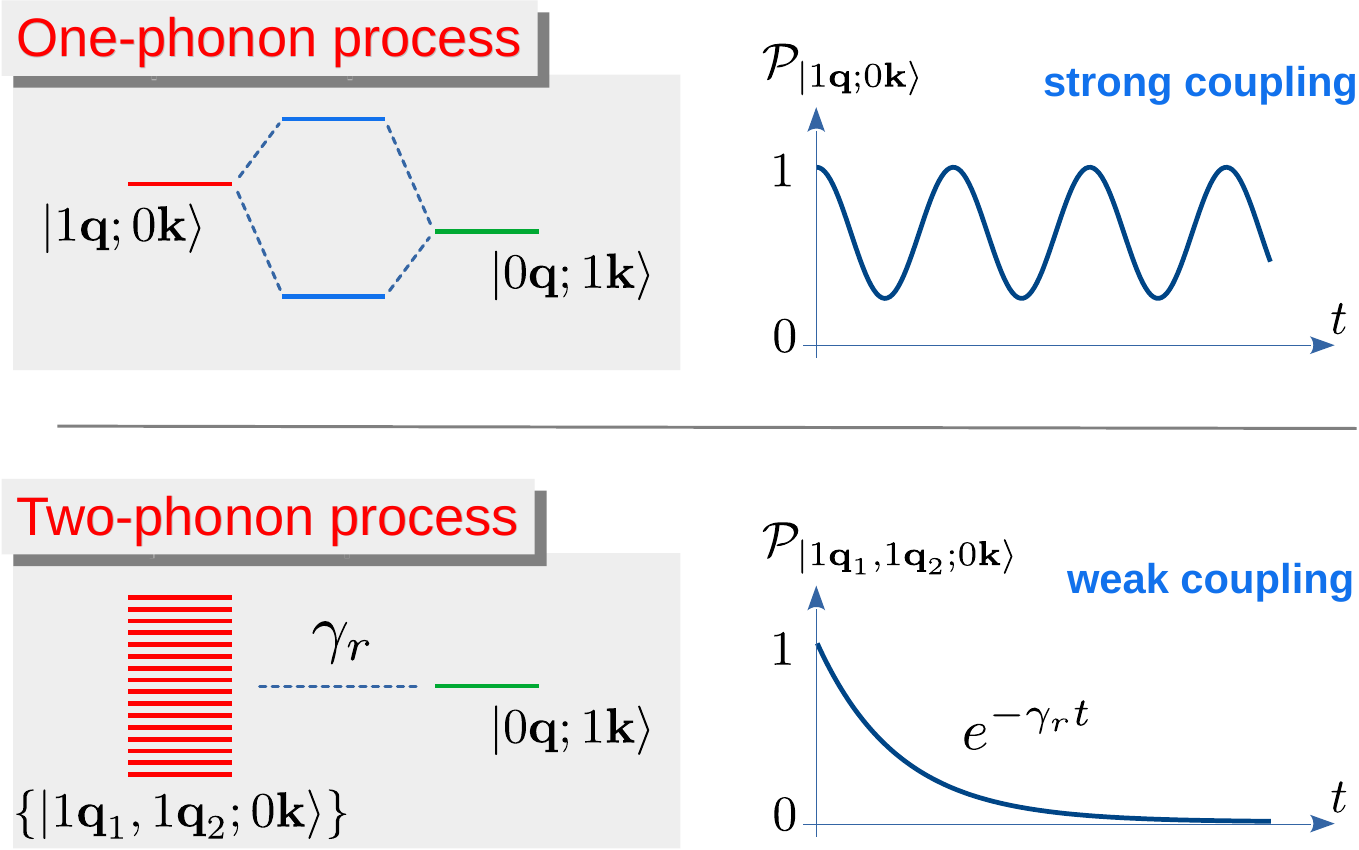} % for an image file named example_figure.*
	% Pick an appropriate width - in print, figures are usually one or two columns wide, which can
	% be approximated by 0.3\textwidth or 0.6\textwidth respectively. Use appropriate label sizes.

	% Captions go below figures
	\caption{\textbf{Strong and weak couplings of phonons to the electromagnetic field in vacuum.} Phonons of wavevector $\textbf{q}$ are coupled to photons of wavevector $\textbf{k}$. (Top) One-phonon process: the selection rule $\textbf{q}=\textbf{k}$ leads to the strong coupling regime and the formation of phonon-polaritons with a reversible energy exchange between phonons and photons, characterized by the oscillating probability $\mathcal{P}_{|\rm{1}\bf{q};\rm{0}\bf{k}\rangle}$ to be in a one-phonon state. (Bottom) Two-phonon process: the selection rule $\textbf{q}_1+\textbf{q}_2=\textbf{k}$ applies to a continuum of two-phonon states, thus leading to the weak-coupling regime and the irreversible radiative decay of two-phonon states, characterized by the exponentially decaying probability $\mathcal{P}_{|\rm{1}\bf{q}_1,\rm{1}\bf{q}_2;\rm{0}\bf{k}\rangle}$ to be in a two-phonon state.}
	\label{fig0} % give each figure a logical label name
\end{figure}
%%%%%%%%%%%%%%%%%%%%%%%%%%%%%%%%%%%%%%%%%%%%%%%%%%%%%%%%%%%%%%
%%%%%%%%%%%%%%%%%%%%%%%%%%%%%%%%%%%%%%%%%%%%%%%%%%%%%%%%%%%%%%
\subsection*{Strong and weak coupling regimes}
Besides their different spectral signatures, one-phonon and two-phonon radiative decays further differ by the regime of the light-matter interaction. In the case of a one-phonon process, the photon-phonon interaction is in the strong coupling regime [Fig.\ref{fig0}, top]. A bulk crystal is translationally invariant in three dimensions as is the electromagnetic field in vacuum. Because of the wavevector conservation rule, photons and optical phonons anticross, giving rise to phonon-polaritons \cite{Car96,Kli07}. There is therefore a reversible energy exchange between phonons and photons, and no possible intrinsic radiative decay, because phonon-polaritons are the eigenstates of the photon-phonon system in the strong coupling regime [Fig.\ref{fig0}, top]. The exact analogous situation happens for the well-known problem of the photon-exciton interaction in direct-gap bulk crystals \cite{Car96,Kli07}. As detailed in Ref.\cite{Bur95}, within the exciton-polariton picture, radiative decay of excitons can only occur through the conversion of polaritons into photons at the surfaces of a perfect crystal. More generally, any interface or defect breaking the translational invariance of the crystal could induce the excitonic radiative decay, meaning that the excitonic PL in bulk semiconductors is an extrinsic process. As a consequence, at low excitation power when one-phonon radiative decay dominates [Fig.\ref{figPLMIR}a], the phononic PL spectrum is also of extrinsic nature. We indeed observe a variety of phononic PL spectra at low excitation power depending on the probed microcrystallite (see Fig.S3a), and we attribute the line at 193 meV in Fig.\ref{figPLMIR}a to a defect producing a localized mode within the reststrahlen band.

In contrast, two-phonon radiative decay is an intrinsic process because the light-matter interaction is in the weak coupling regime [Fig.\ref{fig0}, bottom]. Since the phonon wavevectors $\textbf{q}_1$ and $\textbf{q}_2$ can span the full reciprocal phase provided that $\textbf{q}_1+\textbf{q}_2=\textbf{k}$ where $\textbf{k}$ is the photon wavevector, there is a continuum of two-phonon states coupled to a given photon state. The light-matter interaction is in a continuum-to-discrete state configuration that leads to the weak coupling regime, characterized by an irreversible radiative decay of the two-phonon states. As a matter of fact, at high excitation power when two-phonon radiative decay dominates [Fig.\ref{figPLMIR}a], the phononic PL spectrum is of intrinsic nature. It is expected to follow the spectral dependence of the two-phonon density of states, in a first approximation assuming constant matrix elements as a function of the phonon pair states. \textit{Ab initio} calculations of the r-BN bandstructure provide a reasonable agreement with our experimental results (see Fig.S3b and details in Methods), further confirming the evidence of intrinsic two-phonon radiative decay.

We highlight that intrinsic radiative decay in the weak coupling regime is predicted in the opposite limit of thin layers, for one-phonon processes \cite{Cas22PRX,Cas22PRR}. This range of BN thickness was recently investigated in Ref.\cite{abo25,guo25} reporting electroluminescence of hyperbolic phonon-polaritons in h-BN by means of an extrinsic emission process where these non-radiative hybrid excitations of large wavevectors lying outside the light cone are elastically scattered by making use of extended defects \cite{abo25} or gold nanodisks \cite{guo25} in order to accomodate the momentum mismatch. The search for intrinsic spontaneous emission of light by radiative 2D phonons lying inside the light cone appears as an exciting perspective, in a complementary way to the intrinsic two-phonon radiative decay demonstrated here.
%%%%%%%%%%%%%%%%%%%%%%%%%%%%%%%%%%%%%%%%%%%%%%%%%%%%%%%%%%%%%%
\subsection*{Concluding remarks}

The observation of spontaneous emission of light by non-equilibrium phonons opens fascinating perspectives in different fields. The phonon bath appears as an untapped resource for implementing novel schemes for the light-matter interaction and cavity quantum electrodynamics experiments at the solid-state. Incidentally, the non-thermal radiation from phonons breaks down the assumption of local thermodynamic equilibrium at the basis of the local Kirchhoff law and the reciprocity relation between absorptivity and emissivity in generalized descriptions of radiating bodies \cite{greffet}. Moreover, the evidence of phononic PL paves the way for novel strategies in infrared light-emitting devices. The optical phonon energy being strongly material-dependent, leveraging the non-thermal emission from phonons in a great variety of polar crystals would allow spanning both mid- and far-infrared domains \cite{Cas22PRR}. Furthermore, the possibility of bypassing heat generation by the luminescence of non-equilibrium phonons could be the basis of disruptive strategies in thermal management.
%%%%%%%%%%%%%%%%%%%%%%%%%%%%%%%%%%%%%%%%%%%%%%%%%%%%%%%%%%%%%%
\subsection*{Methods}
\subsubsection*{Optical setup}
PL spectroscopy is performed by means of a home-made cryogenic scanning confocal microscope, with an achromatic design allowing ultrabroadband optical spectroscopy with a spatial resolution given by the diffraction limit until the UV-C spectral range \cite{Val19,Rou21,Rou22,Moo25}. Two-photon excitation is implemented with the second harmonic of a continuous-wave mode-locked Ti:sapphire oscillator delivering trains of 140-fs pulses at an 80-MHz repetition rate. The excitation laser beam is focused to a diffraction-limited spot by a Schwarzschild objective (0.5-numerical aperture) located inside the closed-cycle cryostat equipped with CaF$_2$ optical windows. The emission is collected by means of an achromatic optical system comprising a pinhole for confocal filtering. For detection of the excitonic PL in the UV-C range, the signal is dispersed in a Czerny–Turner spectrometer with a 500-mm focal length and a 1,200 grooves.mm$^{–1}$ ruled grating, and finally detected by a back-illuminated charge-coupled device camera with a 13.5-$\mu$m pixel size. For detection of the phononic PL in the MIR range, the signal goes through a Fourier-transform InfraRed (FTIR) spectrometer (Invenio, Bruker\textcopyright) equipped with a mercury-cadmium-telluride (MCT) detector, coupled to a lock-in amplifier synchronized with the mechanical chopper modulating the intensity of the excitation laser beam, for acquisition of interferograms in the step-scan mode. The MIR spectra are corrected from the instrumental response function recorded with a calibrated Globar source.

The control experiments at $T_s$=250 K without any optical excitation ($P$=0) are performed by placing the mechanical chopper in front of the optical window of our cryostat. In this configuration, the output of the lock-in amplifier is non-zero because of the black-body radiation of the chopper itself, but no additional MIR signal from the BN sample surface at $T_s$=250 K is detected, as carefully checked by changing the focus of the microscope objective.
%%%%%%%%%%%%%%%%%%%
\subsubsection*{Boron nitride samples}
Boron nitride microcrystallites with sp$^2$-hybridization of the electronic states are grown by the iron flux method at atmospheric pressure \cite{edg21}. Boron and nitrogen are dissolved at 1550\textcelsius\, in a molten iron ingot, and BN crystals are precipitated on the surface of the iron flux during the cooling process, as detailed in Ref.\cite{des25}. Crystal growth occurs with the honey-comb basal planes stacked preferentially parallel to the surface of the molten flux. The polycrystalline sample is made of crystallites with a lateral size of several tens of $\mu$m at the surface of the shell, a thickness up to several hundreds of $\mu$m, and a section ranging from regular triangles to regular hexagons with intermediate irregular polygonal shapes \cite{des25}. A statistical analysis in tens of microcrystallites by combined micro-Raman and micro-PL experiments did not evidence any correlation between the microcrystallite shape and the crystallographic phase which appears randomly distributed, from one microcrystallite to another, between the most common AA'-stacking of BN leading to h-BN \cite{wat04}, and the ABC-stacking corresponding to r-BN.
%%%%%%%%%%%%%%%%%%%%%%%%
\subsubsection*{\textit{Ab initio} calculations of the two-phonon density of states}
We simulate the phonons of rhombohedral Boron Nitride in density functional theory (DFT) with Quantum ESPRESSO~\cite{Giannozzi2009,Giannozzi2017} and phonopy~\cite{Togo2023}, using norm-conserving semirelativistic pseudopotentials with Perdew-Burke-Ernzerhof~\cite{Perdew1996a} functionals from the PseudoDojo~\cite{vanSetten2018}. We use the recommended 84 Ry Energy cutoff and DFT-D3 van der Waals corrections~\cite{Grimme2010}. Starting from the experimental rhombohedral primitive unit-cell, we relaxed the structure using default convergence thresholds and a $20\times20\times13$ k-points grid.  $3\times3\times3$ supercells with $6\times6\times4$ k-points grid are used to compute forces with displaced atoms. Dielectric and Born effective charge tensors were obtained in Quantum ESPRESSO.

The phonons are then interpolated on a $48\times48\times24$ grid and the one- and two-phonon density of states are computed on 5000 $\omega$ values between $1000$ and $3500$ cm$^{-1}$.  They are obtained as the imaginary part of Lorentzian functions:
\begin{eqnarray}
g_1(\omega) &= \rm{Im} \left(\sum_{\mathbf{q}, \mu} \frac{1}{\omega - \omega_{\mu} - i \eta}\right) \\
g_2(\omega) &= \rm{Im}\left(\sum_{\mathbf{q}, \mu, \nu} \frac{1}{\omega - (\omega_{\mu}+\omega_{\nu} + i \eta)}\right)
\end{eqnarray}
%%%
where $\mathbf{q}$ are phonon momenta on the grid, $\mu,\nu$ are phonon mode indices, and $\eta = 10$ cm$^{-1}$ is chosen arbitrarily to roughly match the peak width of experimental spectra while being much larger than the difference between consecutive $\omega$ values ($0.5$ cm$^{-1}$).
Note that we do not bother with prefactors in the above equations. The densities of states are simply normalized to their largest values.

The two-phonon density of states qualitatively matches the experimental spectrum, except for an extra peak slightly below 2000 cm$^{-1}$ [Fig.S3b]. In practice the quantity probed by the measurements in this region would be the two-phonon density of states weighted by some matrix elements representing two-phonon radiative decay. The suppression of this peak would then be explained by much smaller matrix elements for the two-phonon decay processes involved in this particular peak.
%%%%%%%%%%%%%%%% REFERENCES %%%%%%%%%%%%%%%

\clearpage % Clear all remaining figures and tables then start a new page

% The list of references goes after the main text and before the acknowledgements
% When preparing an initial submission, we recommend you use BibTeX, like this:
%
%\bibliography{science_template} % for a file named science_template.bib

\begin{thebibliography}{1}
\bibitem{sr}Shockley, W. \& Read, W. T. Statistics of the Recombinations of Holes and Electrons. \textit{Phys. Rev.} \textbf{87}, 835 (1952).
\bibitem{h}Hall, R. N. Electron-Hole Recombination in Germanium. \textit{Phys. Rev.} \textbf{87}, 387 (1952).
\bibitem{Bro85}\textit{Nonequilibrium phonon dynamics}, edited by Bron, W. E. NATO ASI Series (Series B: Physics), Vol. 124 (Plenum Press, New York, 1985).
\bibitem{Mac20}Maccabe, G. S. et al. Nano-acoustic resonator with ultralong phonon lifetime. \textit{Science} \textbf{370}, 840 (2020).
\bibitem{Cas22PRX}Cassabois, G. et al. Exciton and Phonon Radiative Linewidths in Monolayer Boron Nitride. \textit{Phys. Rev. X} \textbf{12}, 011057 (2022).

\bibitem{Cas22PRR}Cassabois, G. et al. Superradiance of optical phonons in two-dimensional materials. \textit{Phys. Rev. Res.} \textbf{4}, L032040 (2022).
\bibitem{Ma22}Ma, E. Y. et al. The Reststrahlen Effect in the Optically Thin Limit: A Framework for Resonant Response in Thin Media. \textit{Nano Lett.} \textbf{22}, 8389 (2022).
\bibitem{CassaboisNP}Cassabois, G., Valvin, P. \& Gil, B. Hexagonal boron nitride is an indirect bandgap semiconductor. \textit{Nat. Phot.} \textbf{10}, 262 (2016).
\bibitem{wat11}Watanabe, K. et al. Hexagonal boron nitride as a new ultraviolet luminescent material and its application—Fluorescence properties of hBN single-crystal powder. \textit{Diamond Relat. Mater.} \textbf{20}, 849 (2011).
\bibitem{casPRB16}Cassabois, G. et al. Intervalley scattering in hexagonal boron nitride. \textit{Phys. Rev. B} \textbf{93}, 035207 (2016).
\bibitem{kan21}Kang, T. et al. Ultrafast nonlinear phonon response of few-layer hexagonal boron nitride. \textit{Phys. Rev. B} \textbf{104}, L140302 (2021).
\bibitem{sha23}Sharma, S. et al. Auger Recombination Kinetics of the Free Carriers in Hexagonal Boron Nitride. \textit{ACS Photonics} \textbf{10}, 3586 (2023).
\bibitem{sca25}\v{S}\v{c}ajev, P. et al. Intralayer Carrier Diffusion and Exciton‐Exciton Annihilation in Hexagonal Boron Nitride Investigated by Two‐Color Pump‐Probe Experiments. \textit{Adv. Optical Mater.} \textbf{13}, e01337 (2025).
\bibitem{hut25}Hutchins, W. et al. Ultrafast evanescent heat transfer across solid interfaces via hyperbolic phonon–polariton modes in hexagonal boron nitride. \textit{Nat. Mater.} \textbf{24}, 698 (2025).
\bibitem{Val19}Valvin, P. et al. Deep ultraviolet hyperspectral cryomicroscopy in boron nitride: Photoluminescence in crystals with an ultra-low defect density. \textit{AIP Adv.} \textbf{10}, 075025 (2020).
\bibitem{Rou21}Rousseau, A. et al. Monolayer Boron Nitride: Hyperspectral Imaging in the Deep Ultraviolet. \textit{Nano. Lett.} \textbf{21}, 10133 (2021).
\bibitem{Rou22}Rousseau, A. et al. Bernal Boron Nitride Crystals Identified by Deep-Ultraviolet Cryomicroscopy. \textit{ACS Nano} \textbf{16}, 2756 (2022).
\bibitem{Moo25}Moon, S. et al. Wafer-scale AA-stacked hexagonal boron nitride grown on a GaN substrate. \textit{Nat. Mater.} \textbf{24}, 843 (2025).
\bibitem{gil}Gil, B. et al. Polytypes of sp2-Bonded Boron Nitride. \textit{Crystals} \textbf{12}, 782 (2022).
\bibitem{wat04}Watanabe, K. et al. Direct-bandgap properties and evidence for ultraviolet lasing of hexagonal boron nitride single crystal. \textit{Nat. Mater.} \textbf{3}, 404 (2004).
\bibitem{sato}Sato, T. Influence of Monovalent Anions on the Formation of Rhombohedrat Boron Nitride, rBN. \textit{Proc. Jpn. Acad. Ser. B} \textbf{61}, 459 (1985).
\bibitem{des25}Desrat, W. et al. Growth of rhombohedral boron nitride crystals using an iron flux. \textit{Phys. Rev. Mat.} \textbf{9}, 124001 (2025).
\bibitem{cal19}Caldwell, J. et al. Photonics with hexagonal boron nitride. \textit{Nat. Rev. Mater.} \textbf{4}, 552 (2019).
\bibitem{vuo17}Vuong, T. Q. P. et al. Overtones of interlayer shear modes in the phonon-assisted emission spectrum of hexagonal boron nitride. \textit{Phys. Rev. B} \textbf{95}, 045207 (2017).
%\bibitem{eli19}C. Elias \textit{et al.}, Nat. Comm. \textbf{10}, 2639 (2019).
%\bibitem{mak10}K. F. Mak \textit{et al.}, Phys. Rev. Lett. \textbf{105}, 136805 (2010).
%\bibitem{spl10}A. Splendiani \textit{et al.}, Nano Lett. \textbf{10}, 1271 (2010).
\bibitem{vuong}Vuong, T. Q. P. et al. Isotope engineering of van der Waals interactions in hexagonal boron nitride. \textit{Nat. Mater.} \textbf{17}, 152 (2018).
\bibitem{sch19}Schu\'{e}, L. et al. Bright Luminescence from Indirect and Strongly Bound Excitons in h-BN. \textit{Phys. Rev. Lett.} \textbf{122}, 067401 (2019).
\bibitem{pla19}Plaud, A. et al. Exciton-exciton annihilation in hBN. \textit{Appl. Phys. Lett.} \textbf{114}, 232103 (2019).
\bibitem{kio13}Kioupakis, E. et al. Temperature and carrier-density dependence of Auger and radiative recombination in nitride optoelectronic devices. \textit{New J. Phys.} \textbf{15}, 125006 (2013).
\bibitem{wang04}Wang, F. et al. Auger recombination of excitons in one-dimensional systems. \textit{Phys. Rev. B} \textbf{70}, 241403(R) (2004).
\bibitem{sun14}Sun, D. et al. Observation of rapid exciton-exciton annihilation in monolayer molybdenum disulfide. \textit{Nano Lett.} \textbf{14}, 5625 (2014).
\bibitem{ngu11}Nguyen, D. T. et al. Elastic Exciton-Exciton Scattering in Photoexcited Carbon Nanotubes. \textit{Phys. Rev. Lett.} \textbf{107}, 127401 (2011).
\bibitem{moo15}Moody, G. et al. Intrinsic homogeneous linewidth and broadening mechanisms of excitons in monolayer transition metal dichalcogenides. \textit{Nat. Comm.} \textbf{6}, 8315 (2015).
\bibitem{vuongPRB}Vuong, T. Q. P. et al. Exciton-phonon interaction in the strong-coupling regime in hexagonal boron nitride. \textit{Phys. Rev. B} \textbf{95}, 201202(R) (2017).
%\bibitem{cusco}R. Cusc\'{o} \textit{et al.}, Phys. Rev. B \textbf{94}, 155435 (2016).
\bibitem{Car96}\textit{Fundamentals of semiconductors}, edited by Yu, P. Y. \& Cardona, M. (Springer-Verlag, Berlin Heidelberg, 1996).
\bibitem{spa73}Sparks, M. \& Sham, L. J. Theory of Multiphonon Absorption in Insulating Crystals. \textit{Phys. Rev. B} \textbf{8}, 3037 (1973).
\bibitem{Kli07}\textit{Semiconductor optics}, edited by Klingshirn, C. F. (Springer-Verlag, Berlin Heidelberg, 2007).
\bibitem{Bur95}Andreani, L. C. in \textit{Confined Electrons and Photons}, edited by Burstein, E. \& Weisbuch, C. NATO ASI Series (Series B: Physics), Vol. 340 (Springer, Boston, MA, 1995).
\bibitem{abo25}Abou-Hamdan, L. et al. Electroluminescence and energy transfer mediated by hyperbolic polaritons. \textit{Nature} \textbf{639}, 909 (2025).
\bibitem{guo25}Guo, Q. et al. Hyperbolic phonon-polariton electroluminescence in 2D heterostructures. \textit{Nature} \textbf{639}, 915 (2025).
\bibitem{greffet}Greffet, J. J. et al. Light Emission by Nonequilibrium Bodies: Local Kirchhoff Law. \textit{Phys. Rev. X} \textbf{8}, 021008 (2018).
\bibitem{edg21}Li, J. et al. Hexagonal Boron Nitride Crystal Growth from Iron, a Single Component Flux. \textit{ACS Nano} \textbf{15}, 7032 (2021).
\bibitem{Giannozzi2009}Giannozzi, P. et al. QUANTUM ESPRESSO: a modular and open-source software project for quantum simulations of materials. \textit{J. Phys.: Condens. Matter} \textbf{21}, 395502 (2009).
\bibitem{Giannozzi2017}Giannozzi, P. et al. Advanced capabilities for materials modelling with Quantum ESPRESSO. \textit{J. Phys.: Condens. Matter} \textbf{29}, 465901 (2017).
\bibitem{Togo2023}Togo, A. et al. Implementation strategies in phonopy and phono3py. \textit{J. Phys.: Condens. Matter} \textbf{35}, 353001 (2023).
\bibitem{Perdew1996a}Perdew, J. P. et al. Generalized Gradient Approximation Made Simple. \textit{Phys. Rev. Lett.} \textbf{78}, 1396 (1997).
\bibitem{vanSetten2018}van Setten, M. J. et al. The PseudoDojo: Training and grading a 85 element optimized norm-conserving pseudopotential table. \textit{Comput. Phys. Comm.} \textbf{226}, 39 (2018).
\bibitem{Grimme2010}Grimme, S. et al. A consistent and accurate ab initio parametrization of density functional dispersion correction (DFT-D) for the 94 elements H-Pu. \textit{J. Chem. Phys.} \textbf{132}, 154104 (2010).

%
%\bibitem{example}
%A.~N. {Author}, An example reference. \emph{Journal of Improbable Research}
%  \textbf{1}, 67 (2020).
%
%\bibitem{example2}
%F.~M. {Surname}, S.~{Author}, A second example. \emph{Interesting Research
%  Letters} \textbf{32}, 897 (2019).
%
%\bibitem{example_preprint}
%P.~{One}, P.~{Two}, P.~{Three}, {An unpublished preprint}. \emph{preprint}
%  (2021), arXiv:2101.12345.
%
\end{thebibliography}
%\bibliographystyle{sciencemag}

% After the paper has completed peer review and been revised ready for acceptance,
% you should comment out the lines above and copy-paste the contents of your .bbl
% file here instead. This will help ensure that our conversion software works correctly.
% Remember to re-run BibTeX first - check the timestamp!
%
% Example of the first three entries copy-pasted from science_template.bbl:
%

%%%%%%%%%%%%%%%% ACKNOWLEDGEMENTS %%%%%%%%%%%%%%%
\paragraph*{Acknowledgments:}
We gratefully acknowledge T. Cohen for his technical support at the mechanics workshop, A. Rousseau, G. Castanier, A. Piazza and M. Mastrangelo for their contributions to the experimental developments, C. Consejo, S. Ruffenach, and F. Teppe for helpful discussions, and V. Jacques for critical reading of the manuscript.
\paragraph*{Funding:}
This work was supported by the French Agence Nationale de la Recherche through the projects Qfoil (ANR-23-QUAC-0003) and QISE (ANR-24-CE96-0001), and by the Institute for Quantum Technologies in Occitanie.
\paragraph*{Author contributions:}
J.~P. and P.~V. performed the optical measurements and analyzed the data. M.~M. and W.~D. grew the rhombohedral boron nitride crystals. The experimental configuration of the optical measurements was designed by P.~V. and G.~C., and optimized with the help of T.~T., A.~V. and C.~S. DFT calculations were performed by J.~B and T.~S. All authors discussed the results and contributed to the writing of the paper.
\paragraph*{Competing interests:}
There are no competing interests to declare.
\paragraph*{Data and materials availability:}
All data are available in the main text or in the supplementary information. The rhombohedral boron nitride microcrystallites used in this study were specifically grown for the purpose of this project and are not commercially available.
%%%%%%%%%%%%%%%% SUPPLEMENT LIST %%%%%%%%%%%%%%%

% List the contents of your Supplementary Materials, including the numbers of any
% supplementary figures, tables, external data files etc. and any references that are
% cited only in the supplement. In this example, refs. 7-8 are cited only in the supplement.
% Fill out your numbers accordingly and delete any lines that aren't applicable.

%\subsection*{Supplementary materials}
%Materials and Methods\\

%%%%%%%%%%%%%%%% END OF MAIN TEXT %%%%%%%%%%%%%%%

\end{document}